\documentstyle[12pt]{article}
\advance \topmargin by -\headheight
\advance \topmargin by -\headsep
\evensidemargin \oddsidemargin
\begin{document}
\begin{center}

{\bf The Universe out of an Elementary Particle ?}
\end{center}
\bigskip
\begin{center}
\bigskip
E. I. Guendelman\footnote{Electronic address: guendel@bgumail.bgu.ac.il} and 
J. Portnoy\footnote{Electronic address: jportnoy@bgumail.bgu.ac.il}

{\it Physics Department, Ben Gurion University, Beer Sheva, Israel}
\end{center}

\bigskip

\begin{abstract}
   We consider a model of an elementary particle as a $2 + 1$
dimensional brane evolving in a $3 + 1$ dimensional space. Introducing
gauge fields that live in the brane as well as normal surface tension
can lead to a stable "elementary particle" configuration. Considering the
possibility of non vanishing vacuum energy inside the bubble leads,
when gravitational effects are considered, to the possibility of a quantum 
decay of such "elementary particle" into an infinite universe.
Some remarkable features of the quantum
 mechanics of this process are discussed, in particular the relation
between possible boundary conditions and the question of instability
towards Universe  formation is analyzed.

\end{abstract}

\pagebreak

{\bf 1.Introduction}

\bigskip
In recent years it has become evident that the interplay between
elementary particle theory and cosmology is a subject of great relevance
for the understanding of the fundamental features of the Early Universe.
In this paper we propose yet another possible connection between
elementary particles and cosmology, which is that the
whole universe may have had as its origin in  the decay of single
elementary particle.
This can be regarded as a modern realization of the vision Lema\^{\i}tre 
\cite{L}
had regarding the origin of the universe in the decay of a
"primeval atom".
In recent years a number of authors \cite{G}
 have studied the possibility that
a small bubble of false vacuum could evolve into a whole universe,
including the quantum mechanical possibility of tunneling towards
an infinite universe.
Such false vacuum bubbles do not qualify as "elementary particles "
since they are not static-classical configurations.
 From a physical point of view it is hard to see how for a
 state which is not at least quasi-stationary, tunneling could be
physically relevant since we do not have then a situation, like
the well known $\alpha$ decay process, where the particle can be 
available for decay for a very long time and although the decay
rate can be very small, the very long time of existence of the
quasi stationary state renders the decay relevant from a physical 
point of view. Here we build just such an scenario.
Furthermore, we shall also find some remarkable features associated
with the quantum mechanics of the tunneling process that can lead
to the formation of a universe in this way. In particular, we discuss
the relation between possible boundary conditions on the wave
function and the question of instability towards universe formation.

\bigskip
\begin{center}
{\bf 2.Stable particle-like membrane solutions
in flat space-time } 
\end{center}
We want to view an elementary particle as a 2 + 1-dimensional 
membrane evolving in a 3 + 1-dimensional embedding space-time.
We associate to this membrane, following the standard approach
 to the theory of extended objects \cite {Y}, a surface tension.
If this was the end of the story, we could not achieve a stable
configurations since the surface tension alone wants to make the
surface of the membrane as small as possible and left without anything
to balance it, it will lead to a collapse of the membrane.
In order to compensate the effect of the surface tension we consider
the effect of matter fields that live in the membrane itself,
for example a 2 + 1-dimensional gauge theory defined on the brane surface.
The simplest form for the membrane action
 that incorporates the above requirements is \cite {EA}:
\begin{equation}
S=\sigma _{0} \int \sqrt{-h} d^{3}y + \lambda \int \sqrt{-h}
F_{\alpha \beta} F^{\alpha \beta} d^{3}y
\end{equation}
where $h = det(h _{\alpha \beta})\  \alpha ,\beta = 0,1,2\  and\ 
h _{\alpha \beta}$ is the induced metric on the surface as 
given by
\begin{equation}
h _{\alpha \beta} = \frac {\partial x^{\mu}(y)}{\partial y^{\alpha}}
\frac {\partial x^{\nu}(y)}{\partial y^{\beta}} g_{\mu \nu} [x(y)] 
\end{equation}
where $g_{\mu \nu} (\mu ,\nu =0,1,2,3)$ is the embedding 3 + 1-dimensional
space time and $x^{\mu }$ is the position of the membrane is 
this embedding space-time, and $F_{\alpha \beta} F^{\alpha \beta} \equiv 
F_{\alpha \beta} F_{\delta  \epsilon} h ^{\alpha \delta} 
h ^{\beta \epsilon}$.
Considering a spherically symmetric bubble and using spherical coordinates
$\theta ,\phi $ , the simplest non-trivial potential that respects
spherical symmetry (up to a gauge transformation) is the magnetic
monopole configuration, given by 
\begin{equation}
A_{\phi} =f (1- cos\theta)
\end{equation}
which means that
\begin{equation}
F_{\theta \phi} =f sin\theta
\end{equation}
Considering the most general spherically symmetric metric 
for the 2 + 1-surface (c=1)
\begin{equation}
ds^{2}=h _{\alpha \beta} dy^{\alpha} dy^{\beta}=-d\tau^{2}+r^{2}(\tau)
(d\theta^{2}+sin^{2}\theta d\phi^{2})
\end{equation}
In this case the form (4) leads to $F_{\alpha \beta} F^{\alpha \beta} = 
\frac {2 f^{2}}{r^{4}} $, which leads to an action of the form
\begin{equation}
S= \lambda \int 8 \pi \frac {f^{2}}{r^{2} (\tau )} d\tau +
4 \pi \int \sigma _{0} r^{2} d\tau
\end{equation}
we can therefore regard the gauge field contribution as an r-dependent
contribution to the surface tension 
\begin{equation}
\sigma = \sigma _{0} + \frac {\sigma_{1}}{r^{4}}
\end{equation}
$\sigma_{1}$  being  $2 \lambda f^{2}$.
The energy of the static wall being $4 \pi r^{2}\sigma$ has 
a non trivial minimum for any $ \lambda > 0 $which permit a stable 
configuration.
In addition to allowing for the stabilization of the bubble,
the 2 + 1-internal gauge fields can also play a role in defining the 
electromagnetic currents that compete to external 3 + 1 gauge 
fields \cite {EA}.

\bigskip
\begin{center}

{\bf 3.The effect of gravity and of an internal false vacuum}

\end{center}

Generically, thin layers define boundaries between different
phases and unless there is a reason, those different phases have different 
values for their energy densities. In what follows, we shall consider the 
case when the internal phase, defined by the inside of the membrane,
has a non vanishing positive vacuum energy.
The resulting system, as we shall see in section 4, display the
necessary features which we are after in this work, that is, the
possibility of defining a metastable, long lived configuration  which
can be identified as an elementary particle, and which has the possibility, 
when gravitational and quantum mechanical effects are taken into
account, of decaying into an infinite universe.
In this case, we can also solve for the dynamics of the membrane
by the W. Israel method \cite{I}, which determine the discontinuity of the
extrinsic curvatures along the thin shell.
The internal solution, which corresponds to the existence of a 
non-vanishing vacuum energy density with a $T^{\mu \nu}=g^{\mu \nu} \rho$,
is the well known de Sitter space defined by
\begin{equation}
ds^{2} = -(1-\chi ^{2} r^{2})dt^{2} + \frac {dr^{2}} {(1-\chi ^{2} r^{2})} +
r^{2}(d\theta ^{2} + sin^{2} \theta d\phi ^{2})
\end{equation}
where $\chi ^{2} \ =\ \frac {8\pi \rho G}{3}$, $\rho$ is the energy density.
Outside, in the empty space, we can have only a Schwarzschild space
time according to Birkhoff's theorem
\begin{equation}
ds^{2}= - (1 - \frac {2GM}{r})dt^{2} + \frac {dr^{2}}{(1 - \frac {2GM}{r})} +
r^{2}(d\theta ^{2} + sin^{2} \theta d\phi ^{2})
\end{equation}
In the boundary, we have a singular energy momentum tensor. In order to 
obtain the Israel conditions, the simplest way is to use the Gaussian
normal coordinates.
This is done as follows: denoting the 2 + 1 dimensional hypersurface
of the wall by $\Sigma$, we begin by introducing a coordinate system in
$\Sigma$. For definiteness, two of the coordinates can be taken to be
the angular variables $\theta$ and  $\phi $, which are always well
defined up to an overall rotation for a spherical symmetric
configuration. For the other coordinate in the wall, one can use the
proper time variable $\tau$ that would be measured by an observer 
moving along with the wall. Now consider geodesics normal to $\Sigma $,
$|\eta |$ is then defined as the proper length along one such geodesics,
starting from the surface $\Sigma $ to a given point outside $\Sigma $ .
 We adopt the convention that $\eta $ is taken to be positive in the
Schwarzschild regime and negative in the de Sitter regime. $\eta = 0 $
is of course the position of the wall. At least in a neighborhood of
the wall, any point is going to be intersected by one and only one
such geodesics. Specifying then the associated value of $\eta $ and
the coordinates of the point in $\Sigma $ where the geodesics originated,
we can define in this way a coordinate system in a neighborhood of $\Sigma $.
Thus the full set of coordinates is given by $x^{\mu} = (x^{i}, \eta),
x^{i} = (\tau, \theta ,\phi)$.
In these coordinates $g^{\eta \eta}=g_{\eta \eta}=1$ and 
$g^{\eta i}=g_{\eta i}=0$. Also, we define $\xi _{\mu}$ to be the normal
to an $\eta$ = constant hypersurface, which in Gaussian normal coordinates
has the simple form $\xi ^{\mu}$ = $\xi _{\mu}$ = (0,0,0,1).
We then define intrinsic curvature to an $\eta$ = constant surface as
\begin{equation}
K_{ij}=\xi_{i;j}=\frac{\partial \xi_{i}}{\partial x_{j}}
-\Gamma ^{\eta}_{ij}= -\Gamma ^{\eta}_{ij}=
\frac {1}{2} \partial _{\eta} g_{ij}
\end{equation}
In terms of these variables, Einstein's equations take the form
(as usual $G_{\mu \nu }\equiv R_{\mu \nu } - \frac {1}{2} g_{\mu \nu } R$)
\begin{equation}
G ^{\eta}_{\eta}=- \frac {1}{2} {}^{(3)} \! R + \frac {1}{2} 
[(K^{2})^{i}_{i}-K^{i}_{i} K^{j}_{j}] = 8 \pi G T^{\eta}_{\eta}
\end{equation}
\begin{equation}
G ^{\eta}_{i} = K^{j}_{i|j} - K^{j}_{j|i} = 8 \pi G T^{\eta}_{i}
\end{equation}
\begin{equation}
G ^{i}_{j}= {}^{(3)} \! G ^{i}_{j} + K^{l}_{l} K^{i}_{j} -
\frac {1}{2} \delta^{i}_{j} [(K^{2})^{l}_{l} + (K^{l}_{l})^{2}]+
\partial _{\eta} [K_{j}^{i}-\delta^{i}_{j} K^{l}_{l}]=8 \pi G T^{i}_{j}
\end{equation}
Where $\mid$ means covariant derivative in a three dimensional sense 
(in the 2+1 dimensional space of coordinates $(\tau , \theta ,\phi )$,
which we denote with Latin indices i,j,k,l,m ...). Also quantities 
$ {}^{(3)} \! R $, ${}^{(3)} \! G ^{i}_{j}$, etc. are to be evaluated
as if they concerned to a purely 3-dimensional metric $g_{ij}$, without
any reference as to how it is embedded in the higher four dimensional  space.
By definition, for a thin wall, the energy moment tensor $T^{\mu \nu}$
has a delta function singularity at the wall, so one can define a
surface energy momentum $S^{\mu \nu}$, by writing
\begin{equation}
T^{\mu \nu}=S^{\mu \nu} \delta (\eta) + regular \   terms
\end{equation} 
When the energy momentum tensor (14) is inserted into the field equations
(11-13), we obtain that (11) and (12) are satisfied automatically, provided they
are satisfied for $\eta \neq 0$ ( so that $K_{ij}$ does not acquire a delta
 function singularity). Eq.(13) however, when integrated from 
$\eta=-\epsilon \  to\  \eta=+\epsilon \ (\epsilon \rightarrow 0\  and
\  \epsilon > 0)$,
leads to the discontinuity conditions
\begin{equation}
\gamma _{j}^{i}-\delta^{i}_{j} Tr \gamma = 8 \pi G S^{i}_{j}
\end{equation}
where $\gamma _{ij}$=
$  \lim_{\epsilon \rightarrow 0} $
 $(K_{ij}(\eta = +\epsilon )
-K_{ij}(\eta =- \epsilon ))  $.
Solving from (15) for the trace of $\gamma ^{i}_{j}$ and substituting back
into (15), we get
\begin{equation}
\gamma _{j}^{i}= 8 \pi G[S _{j}^{i}-\frac {1}{2}\delta^{i}_{j} Tr S] 
\end{equation}
It is easy to see that the local conservation of $T_{\mu \nu}$, when
it is of the form (14) implies that
\begin{equation}
S^{\eta \eta} = S^{\eta i} =0
\end{equation}
If we combine (17) with the demand of spherical symmetry, we arrive
at the form
\begin{equation}
S^{\mu \nu}=\sigma (\tau ) u^{\mu} u^{\nu} - \omega (\tau) [h^{\mu \nu} 
+ u^{\mu} u^{\nu}]
\end{equation}
where
\begin{equation}
h^{\mu \nu}=g^{\mu \nu}-\xi^{\mu} \xi^{\nu}
\end{equation}
is the metric projected onto the hypersurface of the wall, and
\begin{equation}
u^{\nu}=(1,0,0,0)=four\  velocity\  of\  the\ wall
\end{equation}
 In (18), $\sigma$ has the interpretation of energy per unit surface, as
detected by an observer at rest with respect to the wall, and $\omega (\tau)$
has the interpretation of surface tension.
The form (18) is also obtained directly from the variation of the matter
Lagrangian (1,4). In this case we obtain $\sigma =\sigma _{0} + \frac
{\sigma _{1}}{r^{4}(\tau )}, \omega = \sigma _{0} -
 \frac {\sigma _{1}}{r^{4}(\tau )}$ in agreement with (7).
We now consider the matching through a thin spherical wall of the
asymptotically flat (as $r \rightarrow \infty $) Schwarzschild solution (9),
to the de Sitter solution (8). First consider the discontinuity of $K_{\theta 
\theta }$. Using equation(10), we get for $K_{\theta \theta }$
\begin{equation}
K _{\theta \theta}=\frac {1}{2} \partial _{\eta} g _{\theta \theta}=
\frac {1}{2} \xi^{\mu} \partial _{\mu}   g _{\theta \theta}
\end{equation}
In the last step, in (21) we have expressed the derivative in the direction
of the normal $\partial _{\eta}$, in an arbitrary coordinate system
using the normal vector $\xi _{\mu}$.
Adopting the convention where the normal points from the de Sitter space
towards the Schwarzschild space, we have, denoting by - quantities
referred to the de Sitter space and + to the related to the Schwarzschild
space, that according to (16), (18),  and  (19),
\begin{equation}
K_{\theta \theta}^{-} - K_{\theta \theta}^{+}=4 \pi G \sigma r^{2}
\end{equation}
Both de Sitter and Schwarzschild can be expressed in the form
\begin{equation}
ds^{2}_{\pm}=-A_{\pm}(r) dt^{2}_{\pm} + A_{\pm}^{-1}(r) dr^{2} +
r^{2}(d\theta ^{2}+sin^{2}\theta d\phi ^{2})
\end{equation}
In these coordinates $U^{\mu}_{\pm}=( \dot {t_{\pm}} , \dot {r},0,0)$
 (for the velocity of the membrane) and the normal to the
membrane  $\xi^{\mu}_{\pm}=(A^{-1}_{\pm} \dot{r},\beta _{\pm},0,0)$,
$\beta _{\pm}= \sqrt{A_{\pm}+\dot{r}^{2}} $.
There is the possibility of + or - sign for the $ \sqrt{A_{\pm}+\dot{r}^{2}}$.
Full information concerning signs is obtained when going to a globally
good coordinate system. Then $K_{\theta \theta}^{+}$ is given by
\begin{equation}
K _{\theta \theta}^{\pm}=
\frac {1}{2} \xi^{\mu} \partial _{\mu} r^{2}=r \beta _{\pm}
\end{equation}

\bigskip
\begin{center}

{\bf 4. The effective potential for the membrane: stable 
"particle-like" solutions and their possibility of
tunneling to an infinite universe}

\end{center}

Using equations (24), (22) and the specific values of $A_{\pm}$ in our case,
we obtain,
\begin{equation}
\sqrt{1-\chi ^{2} r^{2}+\dot {r}^{2}}-\sqrt{1-\frac {2GM}{r}+\dot {r}^{2}}
= 4 \pi G \sigma r
\end{equation}
where
\begin{equation}
\sigma =\sigma_{0} +\frac{\sigma _{1}}{r^{4}}
\end{equation}
After solving for $\sqrt{1-\chi ^{2} r^{2}+\dot {r}^{2}}$ and squaring
both sides and afterwards solving for $\sqrt{1-\frac {2GM}{r}+\dot {r}^{2}}$
and squaring again, we obtain the following equation
\begin{equation}
\dot{r}^{2} + V_{eff}(M,r)=-1 \ , \ V_{eff}=
2\frac{GM}{r}(2\frac{\alpha ^{2}}{\beta ^{2}}-1)-\frac{4G^{2}M^{2}}{\beta ^{2}
r^{4}}-\frac{\alpha ^{4}r^{2}}{\beta ^{2}}
\end{equation}
being
\begin{equation}
\alpha ^{2}(r)=\chi ^{2} + (4 \pi \sigma (r) )^{2}
\end{equation}
and
\begin{equation}
\beta(r) = 8 \pi G \sigma (r)
\end{equation}
which resembles the equation of a particle in a potential.
$V_{eff}$ with the form (7) for $\sigma (r)$ is depicted in Fig 1.
(in Fig. 1 the behavior as $r \rightarrow 0$ is not depicted. One finds
$V_{eff} \rightarrow - \infty$ as $r \rightarrow  0$).
As we can see from Fig 1.,$V_{eff}$ allows for a classically stable
solution at r=$r_{min}$ (defined by $\frac {\partial V_{eff}}{\partial r}
=0$ and $\frac {\partial ^{2} V_{eff}}{\partial r^{2}} > 0$).
Since $V_{eff} \rightarrow - \infty$ as $r \rightarrow  \infty$,
such a solution can be only classically stable but quantum mechanically
can decay to a membrane containing an infinite universe inside
(since $r \rightarrow  \infty$) is expected.
In the next chapter we investigate some interesting features of the
quantum mechanics of such a process.
\bigskip

\begin{center}
{\bf 5. Some quantum mechanical aspects of the dynamics of the
membrane}
\end{center}

For the form (23) of the internal (-) and external (+) metrics it is 
guaranteed that the 2 + 1-dimensional energy momentum tensor of the
membrane is conserved in a 2 + 1-dimensional sense \cite{ER}.
 This makes us
expect that a Hamiltonian formalism which makes reference only to the 2 + 1
dynamical variables and not to the embedding space makes sense. Such is the
"proper time" quantization developed in ref.\cite {B}.
 Following (4), we take as the Hamiltonian of the system the mass of
the system.
Using eq. (25), we obtain for the mass
\begin{equation}
M=\frac{\chi ^{2} r^{3}}{2G} - \frac {(4\pi G)^{2} \sigma ^{2} r^{3}}{2G}
+4 \pi \sigma r^{2} (1-\chi ^{2} r^{2} + \dot{r} ^{2})^{1/2}
\end{equation}
This mass M corresponds also to the conserved zero-component
of the 4-momentum as defined for example in \cite {W} page 168. 
In an arbitrary Lorentz frame the standard particle energy
momentum relation $P^{0} = \sqrt{\vec P {}^{2}+M^{2}}$ holds.
Having obtained the Hamiltonian, all the other classical dynamical variables
can be obtained as was done in \cite {B}. The conjugate momentum p will
be equal to
\begin{equation}
p=\frac{\partial L}{\partial \dot{r}}
\end{equation}
The Lagrangian will be equal to
\begin{equation}
L=\dot {r} \int H \frac{d\dot{r}}{\dot{r}^{2}}
\end{equation}
This give for the conjugate momentum \cite{B}
\begin{equation}
p=\int \frac{\partial H}{\partial \dot{r}} \frac{d\dot{r}}{\dot{r}}
\end{equation}
Using H as before we arrive at the value of p, which is equal to:
\begin{equation}
p=4 \pi \sigma r^{2} arcsinh(\frac{\dot{r}}{\sqrt{1-\chi^{2} r^{2}}})
\end{equation}
for the membrane inside horizon and
\begin{equation}
p=4 \pi \sigma r^{2} arccosh(\frac{\dot{r}}{\sqrt{\chi^{2} r^{2}-1}})
\end{equation}
outside.
An arbitrary function of r can be added in the definition of p. Classically
it corresponds to an additional total derivative of a function of r in
the Lagrangian, while in Quantum Mechanics it corresponds to a redefinition
of the wavefunction $\Psi '=e^{if(r)} \Psi$
That means that the Hamiltonian can be taken as
\begin{equation}
H=\frac{\chi ^{2} r^{3}}{2G} - \frac {(4\pi G)^{2} \sigma ^{2} r^{3}}{2G}
+4 \pi \sigma r^{2} \sqrt{1-\chi ^{2} r^{2}}cosh(\frac{p}{4 \pi \sigma r^{2}})
\end{equation}
inside the horizon and
\begin{equation}
H=\frac{\chi ^{2} r^{3}}{2G} - \frac {(4\pi G)^{2} \sigma ^{2} r^{3}}{2G}
+4 \pi \sigma r^{2} \sqrt{\chi ^{2} r^{2}-1}sinh(\frac{p}{4 \pi \sigma r^{2}})
\end{equation}
outside it.
In order to achieve a quantum mechanical approach we shall assume that
\begin{equation}
p=-i\frac{\partial}{\partial r}
\end{equation}
and from this
\begin{equation}
e^{-ia\frac{\partial}{\partial r}}\Psi (r)=\Psi(r-ia)
\end{equation}
and from this assumption, the Schroedinger equation is
\begin{equation}
H\Psi=m\Psi
\end{equation}
being  m the mass parameter of the external Schwarzschild  solution.
Defining the dimensionless variable (in units where $\hbar$ = c = 1)
\begin{equation}
x=\frac{4\pi r^{3}\sigma _{0}}{3}-4 \pi \frac{\sigma_{1}}{r}
\end{equation}
we receive the following difference equation for $\Psi$, interpreting the
order of operators in $\frac{p}{4 \pi \sigma r^{2}}$ as  
$\frac{1}{4 \pi \sigma r^{2}}p$ 
\begin{equation}
f(x)\Psi(x)+g(x)[\Psi (x+i) + \Psi (x-i)]=0
\end{equation}
f and g being real functions of x, inside the horizon, and
\begin{equation}
f(x)\Psi(x)+g(x)[\Psi (x+i) - \Psi (x-i)]=0
\end{equation}
outside.
Expanding  the equation for $\Psi$ outside the horizon, taking $x>>1$
(setting $r\ \sim \frac{1}{\chi}$ and $\chi\ \sim \ G^{1/2}\ \rho _{0}^{1/2}$,
we see that $x>>1$ is satisfied if the typical energy scales determining
$\sigma _{0},\ \sigma _{1}\ and \ \rho _{0}$ are $<<$ Planck scale) 
and keeping the first non vanishing contribution only,
 we receive the equation:
\begin{equation}
-\frac{f}{2g} \Psi (x)=i\frac{\partial \Psi}{\partial x}
\end{equation}
This has the form of a Schroedinger equation (x is time-like outside
the horizon). The solution is:
\begin{equation}
\Psi = Ce^{i \int (-\frac{f}{2g} ) dx}
\end{equation}
where C=constant.
That means that once a bubble passes the horizon it will expand indefinitely,
since from (45) we obtain that $|\Psi|^{2}$ = constant and therefore 
the modulus of the amplitude for the bubble being at 
$r=\frac{1}{\chi} +\epsilon$ ($\epsilon > 0 $ is very small) is the same
as the amplitude for the membrane being at r $\rightarrow \ \infty$ with                                                    
probability equal 1.
Notice that we could in principle avoid the outflow outside
$r=\frac{1}{\chi}$ by taking $\Psi$ = 0 at the point $r=\frac{1}{\chi}$ .
In this case, since $\frac {\partial |\Psi|}{\partial r} =0$
for $r>\frac{1}{\chi}$, this will imply that $\Psi$ = 0 for all
$r>\frac{1}{\chi}$ as well.
Such particles would be stable against decay towards a
 r $\rightarrow \ \infty$   state at this level of approximation.
If we were to expose the particle so defined to an external field 
which can change the vanishing value of the wave function towards
a non zero value at $r=\frac{1}{\chi}$, we would then find of course
that an irreversible flow towards r $\rightarrow \ \infty$ will be 
generated this way.
It is important to notice that the decay of the particle
is an entirely gravitational effect. Indeed when $G \rightarrow 0$
the form of the effective potential contains a minimum at 
$ r= r_{min}$ and as $ r\rightarrow  0$ and $ r \rightarrow \infty$,
$V_{eff} \rightarrow \infty$ leaving no other possibility of
a totally stable discrete spectrum on very general grounds.

\begin{center}
 {\bf 6. Discussion and Prospects for Future Research}
\end{center}
\bigskip
We want to enlarge the research in the future 
by studying the following items:\\
1. In the above discussion, we took the mass as the Hamiltonian.
By ADM theorems this must be. But there is always a possibility 
that the Hamiltonian will be equal to the mass plus other terms that
are equal to zero, if the constraints of the theory are used.
 These terms will not affect the dynamics of the problem, 
but will change the picture we are working with, in particular,
the phase of the wave function depends on the definition of
the canonically conjugate momentum which is not clear.
 We want to arrive at
the Hamiltonian by the canonical formalism, where the meaning
of the phase of the wave function could be better understood.
Also, we want to show how to obtain our results with an arbitrary
choice of time and not only the proper time that has been used 
here. It should be noticed that the proper time method and the canonical
formalism give the same results in other examples that have 
been investigated \cite{B2}.  \\

2.We also saw that when $r>\frac{1}{\chi}$ the equation of motion
change from a second order equation to an equation with first
derivative. We want to study this more in detail.
In particular, the issues related to the boundary conditions
inside and outside the horizon. Also issues related to
the question of information loss (the external first
order equation requires less boundary conditions than
the internal one). It is very interesting to notice that
as we cross the value $r=\frac {1}{\chi}$ the equation becomes
like the time independent Schroedinger equation with r playing
the role of time and therefore the spontaneous appearance
of a time  variable takes place.
 Further implications of this research on the question
of time in quantum cosmology \cite{BI} should be of considerable interest.

3.The problem of the boundary conditions.
Boundary conditions like $\Psi (r=0)=0$ and
 $\Psi (r=\frac{1}{\chi}) \neq 0$
 will give us a particle that can decay to a
 universe and $\Psi (r=\frac{1}{\chi})= 0$
a totally stable particle. We want to study this in greater details.\\

4.The possibility of a solution generalized with spin, as in
Davidson and Paz \cite{D}, but now including gravitational effects 
and the instabilities discovered here.\\

{\bf FIGURE CAPTIONS}

 Fig1. The potential of the membrane V(r) vs. r

\begin{thebibliography}{99}
\bibitem{L}
A. G. Lema\^{\i}tre, "L'Universe en expansion" Annales de la 
Soci\' et\' e Scientifique de Bruxelles A53 (1933) 51;
translated in General relativity and Gravitation {\bf 29} (1997) 641.
Notice that  Lema\^{\i}tre "atom" cannot be referred however as an
"elementary particle" being of a linear size of the order of 
$10^{14}$ cm.

\bibitem{G}
For a review and further references see:
A. H. Guth in {\it The Birth and Early Evolution of our Universe}
Physical Scripta  Nobel Symposium 79   1991

\bibitem{EA}
A. Davidson and E. I. Guendelman, Phys. Lett. {\bf B 251} (1990) 250

\bibitem{I}
W. Israel, Il Nuovo Cimento {\bf 44B} (1966) 1

\bibitem{W}
S. Weinberg, {\it Gravitation and Cosmology},  
Wiley, New York, 1973
\bibitem{B}
V. A. Berezin Phys. Rev {\bf D55} (1997) 2139

\bibitem{ER}
E. Guendelman, A. Rabinowitz, Gen. Rel. Grav. {\bf 28},  (1996) 117

\bibitem{Y}

Yuval Ne'eman, Elena Eizenberg, {\it Membranes and Other Extendons},
 World Scientific  Singapore 1995

\bibitem{BI}

For a review see J. Butterfield and C. I. Isham gr-qc/9901024.

\bibitem{D}
Davidson, A. and Paz, U. Phys. Lett. {\bf B300} (1993) 234

\bibitem{B2}
Berezin, V. A., Boyarsky, A. M.  and Neronov, A. Yu. Phys. Rev. {\bf D 57} 1118 (1998)

\end{thebibliography}
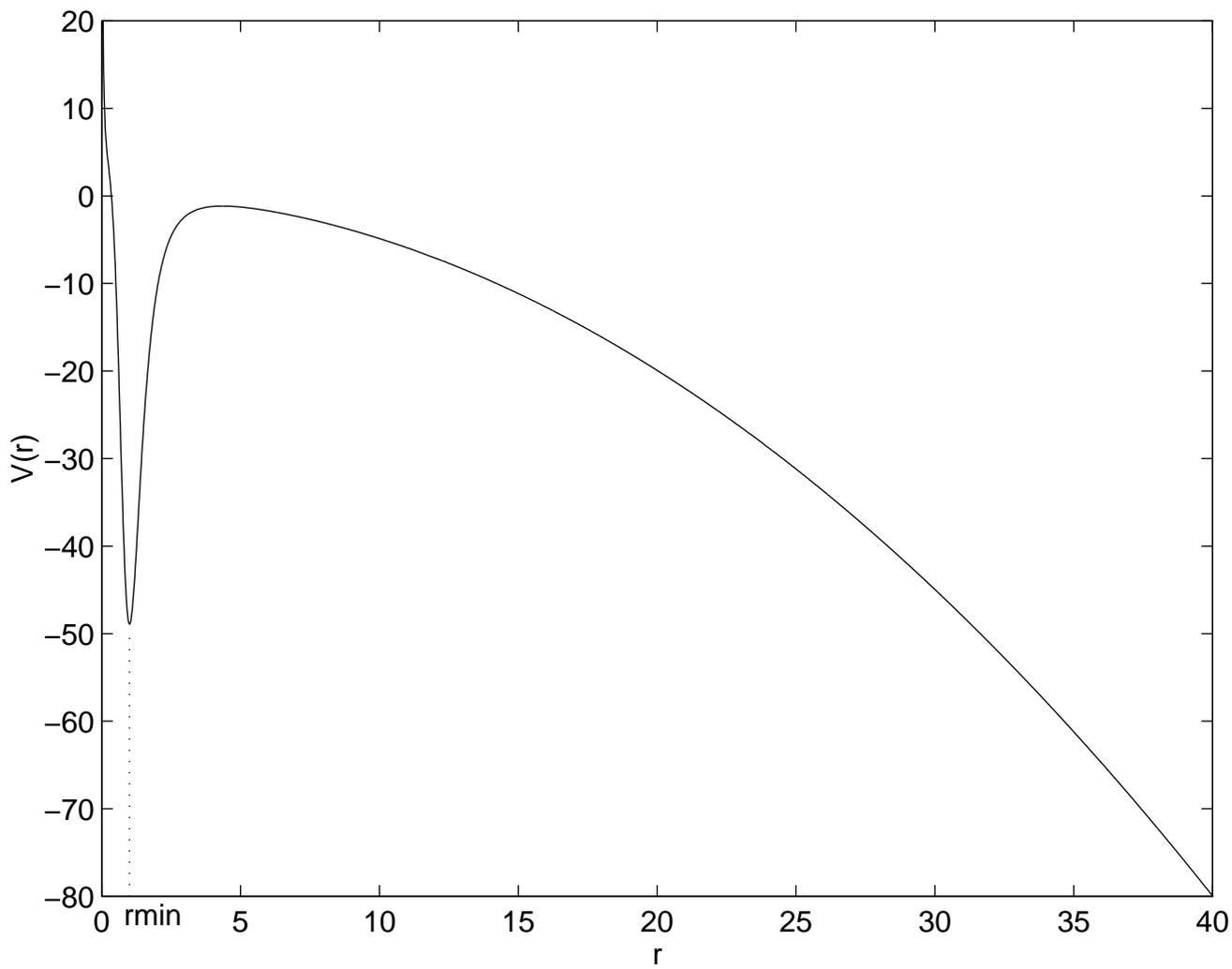
\end{document}